\documentclass[11pt]{article}

\usepackage{slashed}
\usepackage{xcolor}

\def\lsim{\mathrel{\raise.3ex\hbox{$<$\kern-.75em\lower1ex\hbox{$\sim$}}}}
\def\gsim{\mathrel{\raise.3ex\hbox{$>$\kern-.75em\lower1ex\hbox{$\sim$}}}}

\definecolor{red}{rgb}{1.0, 0, 0}
\usepackage{jheppub} 
\usepackage{bm,float}
\usepackage[caption = false]{subfig}
\usepackage[section]{placeins} 
\usepackage{multirow, booktabs, amsthm, xfrac, fix-cm}
\usepackage{url, amsmath, amssymb, color, setspace}

     \def \ebar{\overline{e}} 
 
\def \LC{\epsilon}

\def \d{\text{d}}
\def \PE{\text{PE}}
\def \Nf{N_{\text{f}}}
\def \i{\text{i}}
\def \h{\mathfrak{H}}

\usepackage{etoolbox}
\makeatletter
\patchcmd{\maketitle}{\@fpheader}{ }{ }{ }
\makeatother

\begin{document}

\title{Hilbert Series for Constructing Lagrangians: expanding the
phenomenologist's toolbox}

\author{Landon Lehman}

\author{and Adam Martin}
\affiliation{Department of Physics, University of Notre Dame, Notre Dame, IN
46556}
\emailAdd{llehman@nd.edu}
\emailAdd{amarti41@nd.edu}

\date{\today}
\preprint{ }

\abstract{This note presents the Hilbert series technique to a wider audience
	in the context of constructing group-invariant Lagrangians. This
	technique provides a fast way to calculate the number of operators
	of a specified mass dimension for a given field content, and
	is a useful cross check on more well-known group theoretical methods. In
	addition, at least when restricted to invariants without derivatives,
	the Hilbert series technique supplies a robust way of counting invariants
	in scenarios which, due to the large number of fields involved or to high
	dimensional group representations, are intractable by traditional
	methods. We work out several practical examples.}

\maketitle

\section{Introduction}

In this post-Higgs-discovery era of fundamental physics, phenomenological models
of physics beyond the Standard Model (BSM) are becoming increasingly baroque.
The simplest models have been well-studied and more complex models offer
numerous adjustable parameters that can be tuned to avoid ever more stringent
experimental limits on new physics. One common way to augment the complexity of
a model is to simply add particles transforming under higher-dimensional group
representations. For some representative examples using sizable group
representations in phenomenological settings, see \cite{Ren:2011mh,
Kumericki:2012bh, Earl:2013fpa, Alvarado:2014jva, Cirelli:2005uq} and references
therein.  In addition to complex model building, another line of attack in the
search for new physics takes a general bottom-up approach -- parameterizing BSM
effects through higher dimensional operators formed from SM fields. These
operators are suppressed by powers of some high energy scale, but they become
important for detecting new physics in high precision measurements.

Given a BSM model containing new particle multiplets in different group
representations, one often wants the most general gauge-invariant Lagrangian
containing all of the operators allowed by the postulated symmetries.  The
formation of this Lagrangian allows a full exploration of the model's
experimental signatures. In the simplest cases, with small and familiar group
representations such as $SU(2)$ doublets, it is straight forward to assemble the
Lagrangian almost automatically.  With larger group representations, the
calculation becomes harder, and one often needs to take into account relations
between group invariants (mathematicians refer to these relations as syzygies)
to obtain the correct number of independent terms in the Lagrangian.  A similar
level of calculational difficulty occurs in the task of forming higher
dimensional operators from the Standard Model degrees of freedom.  Even though
the SM group representations are familiar, the high multiplicity of these
representations contained in the operators quickly becomes challenging to deal
with as the operator dimension increases.

A fundamental reason for the prevalence of such group-theoretical calculations
in phenomenological studies is the modern Wilsonian perspective of effective
field theory. From this perspective, the gauge and global symmetries of the
proposed field content completely determine the Lagrangian, and in principle one
must include all possible invariant terms in every operator dimension as part of
the resulting QFT.

The mathematical tool known as the Hilbert series is perfectly suited for such
computations.  The Hilbert series (or Molien or Poincar\'{e} function) is a
generating function encoding information about the number of independent group
invariants that can be formed from some set of multiplets in different
representations.  As we explain in this paper, the Hilbert series provides an
easy cross-check on calculations performed using more familiar group theoretical
techniques, allowing one to ensure the correct number of independent terms in
the Lagrangian. This technique is especially convenient when dealing with large
representations or higher dimensional operators, since the Hilbert series
calculation is easily automated using computer algebra programs such as {\sc
Mathematica}. 

The Hilbert series approach has been developed and used extensively in more
formal theoretical settings. For example, it is often used in calculations
involving the operator spectra of supersymmetric gauge theories
\cite{Pouliot:1998yv, Benvenuti:2006qr, Dolan:2007rq, Gray:2008yu,
	Hanany:2008sb, Chen:2011wn}, SUSY theories on D-branes
	\cite{Butti:2007jv, Feng:2007ur, Forcella:2007wk}, and moduli spaces of
	instantons or vortices \cite{Benvenuti:2010pq, Hanany:2012dm,
	Rodriguez-Gomez:2013dpa, Dey:2013fea, Hanany:2014hia}.  Other references
	from a more mathematical point of view include \cite{Begin:1998hn,
	Hanany:2014dia}.
On the more phenomenological side, Hilbert series methods have been used to
calculate the number of independent flavor invariants in the Standard Model and
various extensions. The Hilbert series for leptonic flavor invariants in the
Standard Model extended by the dimension-5 Weinberg operator was determined for
two and three generations in \cite{Jenkins:2009dy}.  This paper also obtained
the Hilbert series of flavor invariants for the full type-I seesaw model in the
case of two generations, and the series for quark sector flavor invariants for
both two and three generations.  This paper was followed by
\cite{Hanany:2010vu}, which completed the calculation of the Hilbert series for
the seesaw model with three generations, and also computed the series for quark
flavor invariants with four quark generations.

While the Hilbert series technique has been used in formal studies and in the
more phenomenological setting of calculating flavor invariants, it is not yet
well known and appreciated as a tool that can aid in the construction of general
gauge-invariant Lagrangians. In this paper, we aim to rectify this and to add
the Hilbert series to the toolbox of a wider audience. Specifically, in Section
\ref{intro_example}  we introduce the basic Hilbert series concepts through a
simple example, then present the general framework. Then, in Section
\ref{toy_example}, we work through a complete example to solidify the concepts.
Sections \ref{app_1} and \ref{app_2} show some more complicated examples, and we
discuss incorporating derivatives and equations of motion in Section
\ref{deriv}. In Section \ref{conclusion}, we conclude. Some mathematical
background material is included in Appendix \ref{basics}.

\section{Hilbert series basics: $U(1)$ symmetry}
\label{intro_example}

In this section we introduce the key mathematical ingredients of Hilbert series
using a simple example. More formal introductory material to the Hilbert series
can be found in the literature \cite{Feng:2007ur, Hanany:2008sb, Gray:2008yu,
	Benvenuti:2006qr, Hanany:2014dia, Jenkins:2009dy}.

For our example, consider a single complex scalar field charged under a
$U(1)$ symmetry: $\phi \to e^{i \theta}, \phi^* \to e^{-i \theta}$. The gauge
invariant combinations are $(\phi\phi^*)^n$, and there is exactly one
possibility for each $n$. Writing the set of invariants as a series 
\begin{equation}
	\h = \sum_{n=1}^\infty c_n (\phi \phi^*)^n ,
	\label{}
\end{equation}
where $c_n$ is the number of different invariant possibilities for a
given dimension, we have
\begin{equation}
\h = 1 + (\phi\phi^*) + (\phi\phi^*)^2 + (\phi\phi^*)^3 + \dots
\end{equation}
Formally treating $(\phi\phi^*)$ as numbers less than one\footnote{To clarify
that the objects we are manipulating, e.g. $\phi$, are complex numbers rather
than quantum fields, we will refer to $\phi$ and similar objects as spurions.},
this geometric series can be summed,
\begin{equation}
\h = \frac{1}{1-\phi\phi^*}.
\end{equation}
Let us massage this further; the sum above can be replaced by an integral over
$\theta$, the variable that parameterizes the $U(1)$ transformation:
\begin{equation}
\h = \frac 1 {2\pi} \int\limits_{0}^{2\pi}\frac{\d \theta}{(1-\phi\,
	e^{i\,\theta})(1 - \phi^*\, e^{-i\,\theta})}
\end{equation}
Substituting $z = e^{i\, \theta}$, the $\d\theta$ integral becomes a contour
integral around $|z| = 1$.
\begin{equation}
\h = \frac 1 {2\pi i} \oint_{|z|=1}
\frac{\d z}{z}
\frac{1}{(1-\phi\, z)(1 - \frac{\phi^*}{z})} .
\label{ex_int}
\end{equation}
The second piece of the denominator in Eq.~(\ref{ex_int}) can be replaced by
\begin{align}
 \frac{1}{(1-\phi\, z)(1 - \frac{\phi^*}{z})} &= \exp\Big\{ -\log{(1 - \phi z)}
 - \log{(1-\frac{\phi^*}{z})} \Big\} \nonumber \\
& =  \exp\Big\{ \sum\limits_{r = 1}^{\infty}\frac{(\phi z)^r}{r} +
\sum\limits_{r = 1}^{\infty}\Big(\frac{\phi^*}{z}\Big)^r\frac{1}{r} \Big\}.
\label{ex_pleth}
 \end{align}
 To get a better idea of what's going on, let us expand the LHS of
 Eq.~(\ref{ex_pleth}), again treating $\phi$ and $\phi^*$ as small, complex
 numbers rather than quantum fields. To cubic order in both $\phi$ and $\phi^*$
 \begin{align}
\frac{1}{(1-\phi\, z)(1 - \frac{\phi^*}{z})} & =  
\left(1 + \phi\phi^* + (\phi\phi^*)^2 + (\phi\phi^*)^3 + \cdots \right) 
+ z\, \left(\phi + \phi\,(\phi\phi^*) + \phi\,(\phi\phi^*)^2 + \cdots \right) \nonumber \\
& + z^2\, \left(\phi^2 + \phi^2\,(\phi\phi^*) + \cdots \right) + z^3\, \phi^3 +
\frac{\phi^{*3}}{z^3} 
+ \frac{1}{z^2}\left(\phi^{*2} + \phi^{*2}\,(\phi \phi^*) + \cdots \right) \nonumber \\
& + \frac 1 z \left(\phi^* + \phi^*\, (\phi\phi^*) + \phi^*\, (\phi\phi^*)^2 +
\cdots \right)  .
 \end{align}
The series of $U(1)$ invariants sits in the term with no $z$ factors, hence it
is picked out when we multiply by $1/z$ and perform the contour integral in $\d
z$ (Eq.~(\ref{ex_int})). However, inspection of the expansion of
Eq.~(\ref{ex_pleth}) shows it contains all possible arrangements of $\phi$ and
$\phi^*$; if we wanted to pick out the series of charge $+1$ combinations, we
would simply have to multiply by $\frac{1}{z^2}$ before taking the contour
integral, the charge $-2$ can be accessed by multiplying by $z$, etc.

In picking out a particular charge from Eq.~(\ref{ex_pleth}), we are using two
different mathematical facts. The first, already mentioned, is that
Eq.~(\ref{ex_pleth}) generates all possible combinations of $\phi$ and $\phi^*$,
organized by charge. This exponential form of Eq.~(\ref{ex_pleth}) is an example
of a {\em plethystic exponential} \cite{Benvenuti:2006qr, Feng:2007ur,
	Hanany:2008sb, Gray:2008yu,
Hanany:2014dia}, a generating function of
all symmetric combinations of its argument (see~\ref{app:PE}). The second
mathematical construction we employ is integration over the group volume,
$d\theta \equiv \frac{dz}{i\,z}$ for $U(1)$. Integrated over $d\theta$, terms
containing any non-trivial power of $z \to e^{i \,\theta}$ become integrals
$d\theta\,e^{i \, n\,\theta}$ for some integer $n$ and are therefore zero.
Terms with no powers of $z$ -- the $U(1)$ invariants -- remain and are the Hilbert
series $\h$.

While the series of invariants in the $U(1)$ example above could be found
without the aid of Eqs.~(\ref{ex_int}) and (\ref{ex_pleth}), the power of this
approach lies in its generality. Both the plethystic exponential and the
integral over the group parameter $\theta$ can be extended to sets of spurions
transforming under arbitrary representations of arbitrary compact Lie groups. In
addition to generating all possible combinations of spurions such as $\phi$ and
$\phi^*$, the plethystic exponential keeps track of relations among invariants,
or syzygies (see Ref.~\cite{Jenkins:2009dy} for some simple examples of
relations among invariants). Our $U(1)$ example is too simplistic to see these
syzygies, however we will run into relations among invariants when we consider
more complicated setups as in Sections \ref{app_1} and \ref{app_2}. 

To create the plethystic exponential (PE) for a spurion $A$ transforming with
representation $R$ of a connected Lie group, we take
\begin{equation}
	\text{PE}[A, R] = \exp\Big({\sum\limits_{r = 1}^{\infty}
	\frac{A^r \,\chi_{R}(z_j^r)}{r}}\Big).
\end{equation}
Here $\chi_{R}(z_j)$ is the character of the representation $R$ expanded as a
monomial function of the $j$ complex variables on the Cartan sub-algebra
(equivalent to the group rank).  For example, consider  $A$ to be in the
fundamental representation of $SU(2)$, a rank-1 group. The character for the
fundamental representation is
\begin{equation} 
	\Big(z + \frac{1}{z} \Big), 
\end{equation}
where $z$ is a complex number with modulus one (often called a fugacity). The
argument of the plethystic exponential is then
\begin{equation} 
	\sum_{r = 1}^{\infty} \frac{A^r}{r} \left( z^r + \frac{1}{z^r} \right)
	=
	\sum\limits_{r = 1}^{\infty} \Big( \frac{(A z)^r}{r} +
	\frac{A^r}{z^r\,r}\Big)  = -\log{(1-A/z)} - \log{(1-A\,z)}.
\end{equation}
This can be easily extended to more spurions and representations of different
groups -- all we need is a list of the characters for different representations
of Lie groups. For a brief review of characters and further discussion of the
PE, see Appendix \ref{basics}.

Next, the invariants from the PE are picked out by the fact that the characters
of compact Lie groups form an orthonormal set of basis functions on the Cartan
sub-algebra variables \cite{Fuchs:1997jv}. As such, any function $f$ of the
sub-algebra variables can be expanded in terms of them:
\begin{equation}
f(z_j) = \sum_i c_i\,\chi_i(z_j), \quad \int \d \mu\, \chi_i(z) \, \chi^*_j(z) = \delta_{ij},
\label{eq:char_ortho}
\end{equation}
where the $c_i$ are coefficients. The integration $\d \mu$ in
Eq.~(\ref{eq:char_ortho}) is over the Haar measure -- the volume of the group in
question projected onto the Cartan sub-algebra variables (maximal torus). These
volume elements can be found, for example, in Ref.~\cite{Hanany:2008sb} and are
included in Appendix \ref{haar_measure} for convenience. While the PE contains
all possible tensor products of spurions, the group integration projects out
only the invariant combinations, resulting in the Hilbert series $\h$.

Looking back at our $U(1)$ example, we can rephrase the results in this more
general language. The character for a $U(1)$ representation with charge $Q$ is
$z^Q$, so the PE in Eq.~(\ref{ex_pleth}) is the sum of a representation with
charge $Q=+1$ and representation $Q=-1$. The Haar measure for $U(1)$ is $\d
z/z$, and the character orthogonality relation is the usual Fourier series
orthogonality, as is best seen by setting $z = e^{i\,\theta}$:
\begin{equation}
\int\limits_{\theta=0}^{2\pi} \frac{\d\theta}{2\pi}\, e^{i\,\theta\,(Q - Q')} = \delta(Q - Q').
\end{equation}
Continuing, Eq.~(\ref{ex_int}) can be understood as performing a Fourier
expansion of the PE (Eq.~(\ref{ex_pleth})), multiplying by the trivial
representation (i.e.~1) and integrating over the group. Since the characters are
orthonormal, only the part of the PE expansion that lies in the trivial
representation -- the $U(1)$ invariants -- is projected out. Had we wanted to
project out  a different part of the PE, all we have to do is multiply by the
conjugate character before doing the group integration. For example, to project out
the charge +1 combinations, we need to multiply by $1/z = e^{-i\,\theta}$ before
integrating.

Having shown how the PE and group integration can generate the
Hilbert series of invariants for this simple $U(1)$ example, we now want to
apply this method to the Standard Model (and its extensions). Specifically, we
will take some subset of the fields $Q, u^c, d^c, L, e^c, H$, etc.~of the SM
(plus any extensions) as the spurions, dress them with the characters
appropriate to their groups and representations, then form the PE and do the
contour integrations. The different SM spurions can transform differently, and
some may transform under multiple groups. A separate group integration is
carried out for each group under which the fields transform.   If the contour
integrals can be directly calculated, the resulting Hilbert series will be a
rational function of the input spurions, $\h( Q, u^c, d^c, L, e^c, H,\cdots)$.
This function  $\h( Q, u^c, d^c, L, e^c, H,\cdots)$ can then be Taylor expanded
as a multivariate power series in the spurions. The coefficient of any
particular combination of spurions gives the number of independent group-singlet
operators that can be formed from the fields represented by those spurions. For
example, a term $2\, Q^{\dag}\,Q L^{\dag} L$  in the Hilbert series indicates there
are two independent singlets that can be formed from one $Q$ field, one $L$
field, and their hermitean conjugates. We emphasize that, while the Hilbert
series gives the number of invariants, it does not give the particular index
structure, so that must be worked out separately. Before we dive in, a few
comments are in order:
\begin{itemize}
\item Several of the spurions we want to use are fermionic, meaning they
	represent fermion fields. To properly count invariants including
	fermions, we need to extend the PE to handle antisymmetric spurions.
	This can be done by using the fermionic plethystic exponential
	(PEF)~\cite{Hanany:2014dia}. For a fermionic spurion $A$ in
	representation $R$:
\begin{equation}
	PEF[A,R] = \exp\Big( \sum\limits_{r=1}^{\infty}\frac{(-1)^{r+1} A^r
\chi_R(z_j^r) }{r} \Big),
\end{equation}
where, as before, $\chi_R(z_j)$ is the character for representation $R$ as a
function of the Cartan sub-algebra variables.
\item As we are working with fermions (and, eventually, field-strength tensors)
	we must include the Lorentz group representations for these spurions. At
	first sight, this seems problematic since the orthonormality of the
	group characters (the Peter-Weyl theorem) only holds for compact Lie
	groups.  However, since our purpose is solely to count invariants and
	does not involve any dynamics, we can work in Euclidean space, where the
	Lorentz group is compact: $SO(4) \cong SU(2)_R \times SU(2)_L$. For
	simplicity we will take all fermions to be left-handed objects,
	transforming as $(0, \frac 1 2)$ under $SU(2)_L \times SU(2)_R$;
	hermitian conjugate fermions therefore transform in the right-handed
	$(\frac 1 2,0)$ representation. When considering field strength tensors,
	we will work with the objects $X^{\pm}_{\mu\nu} = X_{\mu\nu} \pm
	\i\,\tilde{X}_{\mu\nu}$, which transform in the $(1,0)$ and $(0,1)$
	representations. 
\item While the PE or PEF generate all possible combinations of spurions,
	these constructs contain no information regarding the equations of motion --
	operator relations that go beyond symmetries. How to include the
	equations of motion, and derivatives in general, lies beyond the scope
	of this paper but will appear in a companion paper~\cite{new}.
	A few comments will be sketched out in the discussion in Section \ref{deriv}.
\end{itemize}

\section{Using the Hilbert series: a toy example}
\label{toy_example}

It is instructive to work through a full example of the Hilbert series technique in
order to demonstrate the general procedure.  For this purpose, consider the
Standard Model left-handed fermion doublet $L$, which is a weak isodoublet and a
color singlet. In this example, we will ignore hypercharge, and we denote the
number of generations as $\Nf$. Since $L$ transforms in the fundamental
representation of both $SU(2)_L$ and $SU(2)_W$, the argument of the plethystic
exponential is 
\begin{equation}
	\Nf \; L \left(x + \frac{1}{x}\right) \left(y+ \frac{1}{y} \right) , 
	\label{}
\end{equation}
where $x$ is the complex variable for $SU(2)_L$ and $y$ is the variable for $SU(2)_W$.
Recall that when carrying out the sum for the $\PE$, we need to include the
factor $(-1)^{r+1}$ since $L$ is fermionic. Explicitly, using the Haar measure
for the $SU(2)$ groups as given in Eq.~(\ref{haar_su2}), the Hilbert series for
the spurion $L$ is 
\begin{equation}
	\h(L) = \frac{1}{(2\pi \i)^2} \oint_{|x|=1} \frac{\d x}{x} (1-x^2)
	\oint_{|y|=1}
	\frac{\d y}{y} (1-y^2)  \exp{ \left[ \Nf \sum_{r=1}^{\infty}
	\frac{(-1)^{r+1}}{r}  L^r \left( x^r + \frac{1}{x^r}\right) \left(y^r
	+ \frac{1}{y^r}\right) \right] } .
	\label{ex_explicit}
\end{equation}
Note that $\Nf$ is not a variable in the $\PE$, but is rather a free parameter.
Expanding the PE gives
\begin{equation}
\begin{aligned}
	\exp{ \left[ \Nf \sum_{r=1}^{\infty} 
	\frac{(-1)^{r+1}}{r}  L^r \left( x^r + \frac{1}{x^r}\right) \left(y^r
	+ \frac{1}{y^r}\right) \right] } & =   \\
	 \left( 1 + \frac{L}{x y}
	\right)^{\Nf} 
	\left( 1 + \frac{L x}{y} \right)^{\Nf} &  \left( 1 + \frac{L y}{x}
	\right)^{\Nf} \left(
	1 + L x y \right)^{\Nf} .
\end{aligned}
\end{equation}

In general, it can be computationally challenging to do the contour integrals for
the Haar integration directly, especially if a large number of fields is under
consideration. In this situation, it is better to first expand the integrand in
a Taylor series in the spurion $L$, and then integrate term-by-term up to the
desired power \cite{Pouliot:1998yv}. The poles of the integrand will now all be
at $x = 0$ and $y = 0$, so the residue calculation is much easier. This method will not
give the complete generating function $\h$, but rather the first few terms in
its series expansion.  This is sufficient for most purposes. However, if some
all-order information about the series is needed, or the asymptotic form of the
coefficients as the expansion variables go to infinity
is desired,\footnote{ To obtain asymptotic information about
generating functions, see for example the methods described in
\cite{Wilf:2006:GEN:1204575}.} it will be necessary to directly do the
integrals and obtain the full functional form of $\h$.

For the specific example of Eq.~(\ref{ex_explicit}), we can illustrate this
expansion method by leaving $\Nf$ unspecified and calculating the first few
terms of the Hilbert series for $L$:
\begin{equation}
	\h (L) = \frac{1}{2} (\Nf^2 - \Nf ) L^2 + \frac{1}{6} (\Nf^4 + 5 \Nf^2 )
	L^4 + \frac{1}{144} (5\Nf^6 - 3 \Nf^5 + 17 \Nf^4 -69 \Nf^3 + 50 \Nf^2 )
	L^6 + \dots
	\label{}
\end{equation}
If $\Nf$ is specified, the Hilbert series can easily be calculated exactly, since the
only poles in the integrand of Eq.~(\ref{ex_explicit}) are at $x = 0$ and $y =
0$. Carrying out the contour integrations for the first few values of $\Nf$ gives
the results displayed in Table \ref{lepton_table}.  

\begin{table}[h]
\begin{center}
\begin{tabular}{cc}
\toprule
{\bf $\Nf$ } & {\bf Hilbert series $\h(L)$} \\
\midrule
1 & $1 + L^4$ \\
\midrule
2 & $1 + L^2 + 6 L^4 + L^6 + L^8 $ \\
\midrule
3 & $1 + 3L^2 + 21 L^4 + 20 L^6 + 21 L^8 + 3 L^{10} + L^{12} $ \\
\midrule
4 & $1 + 6L^2 + 56 L^4 + 126 L^6 + 210 L^8 + 126 L^{10} + 56 L^{12} + 6 L^{14} +
L^{16}$ \\
\bottomrule
\end{tabular}
\caption{The Hilbert series for the first few values of $\Nf$, calculated from
	Eq.~(\ref{ex_explicit}). Note that the resulting polynomials are
	palindromic, as noted in \cite{Jenkins:2009dy}. Also, for a given $\Nf$ there is a maximum dimension an operator can have. }
\label{lepton_table}
\end{center}
\end{table}

The entries in Table \ref{lepton_table} can be understood using standard group
theory. In this discussion, $SU(2)_W$ indices will be explicitly displayed, and
$SU(2)_L$ (Lorentz group) indices will not be displayed, but will be contracted
within parentheses. For $\Nf = 1$, there is no possible $L^2$ operator, since
$\LC_{\alpha \beta} (L^\alpha L^\beta ) = 0$.  At the $L^4$ level, the indices
can be contracted in a single non-zero manner: $\LC_{\alpha \beta} \LC_{\delta
\gamma} (L^\alpha L^\delta)(L^\beta L^\gamma)$.  All other methods of index
contraction for $L^4$ are either automatically zero, or can be written as this
one by using the Schouten identities from the two $SU(2)$ groups. At higher
orders, for example $L^8$, it is always possible to use the freedom given by the
two sets of Schouten identities to rewrite the contractions in a form that is
manifestly identically zero, which is why the Hilbert series terminates at
order $L^4$.

For $\Nf = 2$, there is a single operator of order $L^2$, namely $\LC_{\alpha \beta}
(L_1^\alpha L_2^\beta)$. The fact that there are two distinct fields keeps this
from being identically zero. Similarly, at order $L^4$, the six possible
operators are  
\begin{equation}
\begin{aligned}
	\LC_{\alpha \gamma} \LC_{\beta \delta} (L_1^\alpha L_1^\beta
	)(L_1^\gamma L_1^\delta ), \;
	\LC_{\alpha \gamma} \LC_{\beta \delta} (L_2^\alpha L_2^\beta
	)(L_2^\gamma L_2^\delta ), \;
	\LC_{\alpha \gamma} \LC_{\beta \delta} (L_1^\alpha L_2^\beta
	)(L_2^\gamma L_2^\delta ), \; \\
	\LC_{\alpha \gamma} \LC_{\beta \delta} (L_1^\alpha L_1^\beta
	)(L_1^\gamma L_2^\delta ), \;
	\LC_{\alpha \beta} \LC_{\gamma \delta} (L_1^\alpha L_2^\beta
	)(L_1^\gamma L_2^\delta ), \;
	\LC_{\alpha \gamma} \LC_{\beta \delta} (L_1^\alpha L_2^\beta
	)(L_1^\gamma L_2^\delta ).
\end{aligned}
\label{6_terms}
\end{equation}
As a final illustration, for $\Nf = 3$, the three operators of
order $L^2$ are $\LC_{\alpha \beta} (L_1^\alpha L_2^\beta)$, $\LC_{\alpha \beta}
(L_1^\alpha L_3^\beta)$, and $\LC_{\alpha \beta} (L_2^\alpha L_3^\beta)$.

The Hilbert series method can actually go further and give us the exact flavor
content of the operators as in Eq.~(\ref{6_terms}) directly, by doing something
called ``refining'' the series. To do this, instead of putting a generic spurion
$L$ into the PE and multiplying by $\Nf$ as in Eq.~(\ref{ex_explicit}), we put
in $\Nf$ distinct spurions with different labels. For example, if $\Nf = 3$, use
spurions $L_1$, $L_2$, and $L_3$ in the PE. This is also referred to as a
``multigraded'' Hilbert series. Doing this will in general increase the
complexity of the residues that must be computed, so in some cases it may make
the problem intractable. However, for the series of Eq.~(\ref{ex_explicit}) it
is feasible, at least for low values of $\Nf$. Calculating the multigraded
series for $\Nf = 2$ and picking out the $L^4$ terms replicates the results of
Eq.~(\ref{6_terms}). Of course, refining the series does not give the exact
index contraction structure (since there are multiple equivalent ways to
contract the indices), but it does tell us that, for example, there are two
independent operators with the flavor content $L_1^2 L_2^2$.

Because of the increase in computational power required by refining the Hilbert
series, it may sometimes be advantageous to go in the other direction and
``unrefine'' the series. For example, if the Hilbert series involves both the
lepton doublet $L$ and the quark doublet $Q$, it can be useful to not only lump
all of the flavors together, but also to lump together $L$ and $Q$ by setting
both spurions to a common label $t$ in the PE. This can significantly decrease
the computational time required by a progam like {\sc Mathematica}, and may give
all of the information that is necessary for a specific application of the
Hilbert series method.

The example worked out in this section shows that while the results obtained
from the Hilbert series can indeed be replicated using other techniques, the
calculations are often tedious and it is easy to make mistakes and miscount the
operators. For example, showing that there are 126 independent operators of
order $L^{10}$ for $\Nf = 4$ would be a formidable task by standard methods. By
refining the series, the Hilbert series method provides an easy automatic way of
obtaining the desired operators.

\section{Using Hilbert series for Standard Model effective field theory}
\label{app_1}

The Standard Model effective field theory (SMEFT) consists of the Standard Model
Lagrangian $\mathcal{L}_{SM}$ plus operators with (mass) dimension greater than
four that are invariant under the Standard Model gauge group $SU(3)_C \otimes
SU(2)_W \otimes U(1)_Y$ and contain only Standard Model degrees of freedom. At
dimension 5, there is only one possible operator, the Weinberg neutrino-mass
operator \cite{Weinberg:1979sa}.  Continuing on to dimension 6, a classification
of the available operators was done in \cite{Buchmüller1986621}, and the
reduction to a minimal set of dimension-6 operators was carried out in
\cite{Grzadkowski:2010es}, resulting in a set of 63 operators. More recently,
the construction of the minimal set of dimension-7 operators was also completed,
giving 20 independent operators \cite{Lehman:2014jma}. Any BSM physics can be
matched onto this effective field theory by integrating out the postulated new
heavy particles. In the past few years, much work has been done towards
understanding the structure and use of the SMEFT \cite{Grojean:2013kd,
	Jenkins:2013zja, Jenkins:2013wua, Alonso:2013hga, Elias-Miro:2013gya,
	Elias-Miro:2013mua, Alonso:2014zka, Alonso:2014rga, Trott:2014dma,
Henning:2014wua, Elias-Miro:2014eia}.

Calculating the full set of operators for a given dimension with the Hilbert
series requires including covariant derivatives, field strengths, and the
equations of motion. Some thoughts on these ingredients are included in Section
\ref{deriv}, and we leave an in-depth discussion to future work \cite{new}. For
simplicity, in this section we focus on operators without these complicating
factors. This still provides a good example of the practical applications of the
technique.

\subsection{Dimension-6 baryon number violating operators}
\label{sec:dim6B}

The Hilbert series technique proves to be useful for finding independent SMEFT
operators, even when restricted to a subset of the field content. For example,
consider the set of dimension-6 baryon-number-violating operators.
Ref.~\cite{Grzadkowski:2010es} presented five such operators.  However, it has
been noted that only four of these five operators are independent if the flavor
structure is taken into account \cite{Alonso:2014zka, Abbott:1980zj}. It would
be nice to see such operator relations without doing a detailed calculation, and
indeed using the Hilbert series can bring the dependence betwen different
operator structures to light.

Consider the class of baryon-number violating dimension-6 operators with field
content $QQQL$, where $Q$ is the left-handed quark doublet and $L$ is the
lepton doublet. Taking $\Nf = 3$, the argument of the PE for these fields is
\begin{equation}
	3 Q \left(x + \frac{1}{x}\right) \left(y + \frac{1}{y}\right) \left(z_1 + \frac{z_2}{z_1} +
	\frac{1}{z_2} \right) u^{1/6} + 3 L \left(x + \frac{1}{x} \right)
	\left(y + \frac{1}{y} \right) u^{-1/2} ,
	\label{}
\end{equation}
where $x$ is the variable for $SU(2)_W$, $y$ is the variable for $SU(2)_L$, $u$
is the variable for $U(1)_Y$, and $\{z_1, z_2\}$ are the variables for
$SU(3)_C$.  Calculating the multigraded Hilbert series so that we can see the
form of the operators gives
\begin{equation}
	\h(L,Q) = 1 + 57\; L Q^3 + 4818\; L^2 Q^6 + 162774\; L^3 Q^9 +  \dots  ,
	\label{dim6_multi}
\end{equation}
so we expect 57 independent operators of the form $QQQL$ when flavor structure
is included.

Now consider the specific operator structure
\begin{equation}
	\LC_{\alpha \beta \gamma} \LC_{ij} \LC_{kl} \left( Q^{i\alpha}_p
	Q^{j\beta}_r \right) \left(Q_s^{k \gamma} L^l_t \right) .
	\label{bnv1}
\end{equation}
This is symmetric in the flavor indices $\{p,r\}$, so without even considering
additional symmetries implied by $SU(2)_L$ Fierz identities, the maximum number
of independent flavor permutations is $3\times 3 \times 6 = 54 < 57 $ (since a
$3\times 3$ symmetric matrix has only 6 independent entries). Therefore
the operator structure in Eq.~(\ref{bnv1}) does not capture all of the 57
independent operators that we know exist at this order. Perhaps changing the
$SU(2)_W$ index contraction structure will enable us to capture all of the
operators in a single expression. To this end, consider the following structure
with the $SU(2)_W$ contractions ``offset'' from the $SU(2)_L$ contractions:
\begin{equation}
	\LC_{\alpha \beta \gamma} \LC_{ik} \LC_{jl} \left( Q^{i\alpha}_p
	Q^{j\beta}_r \right) \left(Q_s^{k \gamma} L^l_t \right) .
	\label{dim6_1}
\end{equation}
None of the symmetries are immediately apparent in this form, 
so naively one might think that we can now just multiply out the flavor
possibilities and get $3^4 = 81$ operators with different flavor structures. But
this neglects $SU(2)_L$ Fierz identities (see for example
\cite{Dreiner:2008tw}), the relevant one of which gives
\begin{equation}
	\LC_{\alpha \beta \gamma} \LC_{ik} \LC_{jl} \left(Q^{i\alpha}_p Q^{j\beta}_r \right)
	\left(Q^{k\gamma}_s L^l_t \right) = - \LC_{\alpha \beta \gamma} \LC_{ik} \LC_{jl}
	\left[ \left( Q^{i\alpha}_p Q^{k\gamma}_s \right) \left( Q^{j\beta}_r L^l_t \right) +
		\left( Q^{i\alpha}_p L^l_t \right) \left( Q^{j\beta}_r Q^{k\gamma}_s \right)
	\right] .
	\label{}
\end{equation}
Rearranging the first term on the right-hand side using the $SU(2)_W$ Schouten
identity $\LC_{ij}\LC_{mn} = \LC_{im}\LC_{jn} - \LC_{in}\LC_{jm}$, and
relabelling the second term on the right at the cost of a sign gives
\begin{equation}
	\LC_{\alpha \beta \gamma} \LC_{ik} \LC_{jl} \left(Q^{i\alpha}_p Q^{j\beta}_r \right)
	\left(Q^{k\gamma}_s L^l_t \right) = 
	\LC_{\alpha \beta \gamma} \LC_{ik} \LC_{jl} \left[ 2 \left(Q^{i\alpha}_p
		Q^{j\beta}_s \right) \left( Q^{k\gamma}_r L^l_t \right) - \left(
		Q^{i\alpha}_s Q^{j\beta}_r \right) \left( Q^{k\gamma}_p L^l_t
		\right) \right]  .
	\label{}
\end{equation}
This identity tells us that among the 81 different flavor permutations of the
structure in Eq.~(\ref{dim6_1}), there are $3^3 = 27$ relations among operators
differing only in $Q$ flavor permutations. When all of the quark flavors are
identical $(p = s = r)$, the relation is trivial. Since $(p = s = r)$ occurs 3
times for $\Nf = 3$ as in the Standard Model, there are $27 - 3 = 24$ linear
relations among the 81 flavor permutations, leaving $81 - 24 = 57$ independent
permutations! Since the Hilbert series in Eq.~(\ref{dim6_multi}) revealed that
there are only 57 independent operators, we are done; the structure in
Eq.~(\ref{dim6_1}) contains all of the possibilities, and we do not need to
write down a different fermion current structure or different $SU(2)_W$
index contractions.

The Hilbert series technique is by no means restricted to invariants of gauge
symmetries; it can deal just as easily with global symmetries. As one example,
we could have included baryon number for all fields, adding a character and
group integration for that global $U(1)$. Had we included baryon number in the
above example, we would find no invariants -- by construction, since our example
concerned baryon number violation. However, had we considered a wider set of
spurions, both baryon number violating and respecting terms in the Hilbert
series would be generated. In this case, integration over baryon number could be
used to project out different subsets, e.g.~only baryon number respecting
operators, operators violating baryon number by one unit, by two units, etc. Of
course, the same technique could be used for finding operators violating lepton
number or $B-L$.

\subsection{The dimension-7 operator $LLL\overline{e}H$}
\label{sec:dim7}

As another example of applying the Hilbert series to the SMEFT, consider the
class of dimension-7 operators with field content $LLL\overline{e}H$
\cite{Lehman:2014jma}. Here $\overline{e}$ is a left-handed field which can also
be written as $e^c$, and $H$ is the Higgs doublet. In the argument of the
plethystic exponential we then have (again with $\Nf = 3$):
\begin{equation}
	3 L \left(x+\frac{1}{x}\right)\left(y+\frac{1}{y}\right)u^{-1/2} + H
	\left(x+\frac{1}{x}\right)u^{1/2} + 
	3 \overline{e} \left(y+\frac{1}{y}\right) u\; ,
	\label{}
\end{equation}
where $x$ is the variable for $SU(2)_W$, $y$ is the variable for $SU(2)_L$, and
$u$ is the variable for $U(1)_Y$. Calculating the unrefined Hilbert series
gives
\begin{equation}
	\h(t) =1 +  3\; t^4 + 57\; t^5 + 171\; t^6 + 6\; t^8 +\; 144t^9 + 1053\; t^{10} + \dots
	\label{}
\end{equation}
Order $t^5$ contains the dimension-7 operators, and calculating the multi-graded
Hilbert series shows that this order contains only the dimension-7 operators of
the form $L^3 \overline{e} H$, so we know that there are 57 operators with
various flavor structures.

One such possible structure is
\begin{equation}
	\LC_{ij}\LC_{mn} \left( \ebar_p L^i_q \right)\left(L^j_r L^m_s \right) H^n .
	\label{struct}
\end{equation}
As in the previous section, the symmetries are not manifest, so we need to use a
Fierz identity:
\begin{equation}
	\LC_{ij}\LC_{mn} \left( \ebar_p L^i_q \right) \left(L^j_r L^m_s \right)
	H^n = - \LC_{ij}\LC_{mn} \left(\ebar_p L^j_q\right) \left( L^m_s
	L^i_r\right) H^n - \LC_{ij}\LC_{mn} \left(\ebar_p L^m_s\right)
	\left(L^i_r L^j_q \right) H^n .
	\label{fierz_1}
\end{equation}
In the first term on the right-hand side of Eq.~(\ref{fierz_1}), we can switch
the $\{i,j\}$ labels at the cost of a sign (and use the fact that $(z_1 z_2) =
(z_2 z_1)$ for anticommuting 2-component spinors). The second term on the
right-hand side can be rewritten in a similar way after applying the $SU(2)_W$
Schouten identity. The final result is
\begin{equation}
	\LC_{ij}\LC_{mn} (\ebar_p L^i_q)(L^j_r L^m_s) H^n = \LC_{ij}\LC_{mn} H^n
	\left[ (\ebar_p L^i_q)(L^j_r L^m_s) + (\ebar_p L^i_s)(L^j_r L^m_q) -
	(\ebar_p L^i_s )(L^j_q L^m_r) \right] .
	\label{}
\end{equation}
In a similar way as in the previous section, this gives 27 relations among the
81 flavor permutations, this time among operators differing in $L$ flavor
permutations. Again only 24 relations are non-trivial, leaving 57 independent
operators. So the structure in Eq.~(\ref{struct}) encapsulates all of the
independent operators of the form $LLL\overline{e}H$.

Now suppose that instead of starting with the structure in Eq.~(\ref{struct}), the
first guess had been instead
\begin{equation}
	\LC_{ij}\LC_{mn} (\ebar_p L^m_q)(L^i_r L^j_s)H^n .
	\label{}
\end{equation}
Without even considering Fierz identities, we can see that this structure is
antisymmetric in $\{r,s\}$, so it contains a maximum of $3\times 3 \times 3 = 27$
independent flavor permutations, since a $3 \times 3$ antisymmetric matrix has
only 3 independent entries. Thus we know that another structure is needed in order to
get the full 57 independent flavor permutations. In this way the Hilbert series allows
a check on the generality of a specific $SU(2)_L$ and $SU(2)_W$ index
contraction structure.

\section{Hilbert Series for BSM}
\label{app_2}

In the phenomenological study of extensions to the Higgs sector, it is often
necessary to write down the most general form of the Higgs potential including
various multiplets of $SU(2)_W$ with potentially different hypercharges. Usually
only renormalizable terms are included in the potential, but in some cases
higher-dimensional terms are also necessary. In either case, this exercise is a
perfect candidate for using the Hilbert series to check that the Lagrangian
includes a complete set of operators up to a given dimension.

As an explicit example, consider extending the SM Higgs sector by adding a
scalar multiplet $\chi$ which is a quadruplet under $SU(2)_W$ and has
hypercharge $-1/2$, as was done in \cite{Ren:2011mh}. The dimension-2 terms for
the Higgs sector potential are then trivial to construct, and there are no
possible dimension-3 terms, but the task becomes more complicated at dimension
4. Using the character function for the $SU(2)$ quadruplet as given in Appendix
\ref{basics} (Table \ref{character_table}), the argument of the PE for the SM
Higgs $\Phi$ and the new field $\chi$ is
\begin{equation}
	\chi \left(z^3 + z + \frac{1}{z} + \frac{1}{z^3} \right)u^{-1/2} +
	\overline{\chi} \left(z^3 + z + \frac{1}{z} + \frac{1}{z^3} \right)
	u^{1/2} + \Phi \left( z + \frac{1}{z} \right)u^{1/2} + \overline{\Phi}
	\left( z + \frac{1}{z} \right) u^{-1/2} .
	\label{}
\end{equation}
Generating the first few terms of the unrefined Hilbert series gives 
\begin{equation}
	\h(t) = 1 +  2\; t^2 + 11\; t^4 + 31\; t^6 + 94\; t^8 + 222\; t^{10} + \dots  ,
	\label{}
\end{equation}
so we see that there are 11 independent operators at the dimension-4
level.\footnote{As all operators are scalars, counting dimensions is the same as
	counting the number of spurions. This can be generalized by weighting
	each spurion by its mass dimension, i.e $H \to \epsilon H, Q \to
	\epsilon^{3/2}Q$, then collecting like terms.}

As a side note, the field and group representation content for this example is
simple enough that the closed form for the unrefined Hilbert series can be
calculated. The result is
\begin{equation}
	\h (t)  = \frac{ 1 + 4t^4 + 9t^6 + 17t^8 + 13t^{10} + 17t^{12} + 9t^{14}
	+ 4t^{16} + t^{20} }{ (t^4 + t^2 + 1 )^2 (t^2 + 1)^4 (t^2 - 1)^8 }  .
	\label{sseries}
\end{equation}  
Since the complete Hilbert series can be calculated, and since the result is a
rational function of $t$, it is always possible to calculate a closed form for
the coefficients of the series (see for example Chapter 7 of
\cite{Graham:1994:CMF:562056}) . One way to do this is to use the {\sc
Mathematica} function \texttt{SeriesCoefficient}, putting in a general integer
$n$ as the argument for the coefficient order. The resulting expression for the
coefficients of Eq.~(\ref{sseries}) is complicated and not that useful, since
calculating the Taylor series to the necessary order is simple in this case.
However, there might be situations where calculating a closed form for the
coefficients could be useful for figuring out general information about the form
of the series such as the asymptotic form of the coefficients.

Returning to the calculation of the (renormalizable) Higgs potential, by
using {\sc Mathematica} or a similar program, we can generate the
multi-graded Hilbert series and extract the relevant terms, which gives
operators with the following field content:
\begin{equation}
\begin{gathered}
	2\; \chi \chi^\dagger \chi \chi^\dagger, \quad  
	\Phi^\dagger \chi \chi^\dagger \chi^\dagger, \quad 
	\Phi^\dagger \Phi^\dagger \chi^\dagger \chi^\dagger, \quad
	\Phi \chi \chi \chi^\dagger, \quad
	2\; \Phi \Phi^\dagger \chi \chi^\dagger, \quad \\
	\Phi \Phi^\dagger \Phi^\dagger \chi^\dagger, \quad 
	\Phi \Phi \chi \chi, \quad
	\Phi \Phi \Phi^\dagger \chi, \quad
	\Phi \Phi \Phi^\dagger \Phi^\dagger  .
\end{gathered}
\label{potential}
\end{equation}
Counting up the terms we do indeed get eleven. Note that for the field content
$\chi \chi^\dagger \chi \chi^\dagger$, the Hilbert series tells us that there
are two independent terms, which is indeed the conclusion reached in the
paragraph following Eq.~(13) of Ref.~\cite{Ren:2011mh}. The same conclusion
holds for the field content $\Phi^\dagger \Phi \chi^\dagger \chi$. The Hilbert
series provides these results without the need for any detailed calculations
involving group invariants. These calculations can of course be done without
much trouble for these dimension-4 operators, but it is nice to have this check.
Also note that the Hilbert series result (\ref{potential}) gives an operator and
its hermitian conjugate that are not included in the result for the most general
renormalizable Higgs potential given in Eq.~(12) of Ref.~\cite{Ren:2011mh},
namely the operator with field content $\chi \chi \chi^\dagger \Phi$. Working
out the correct $SU(2)_W$ index contractions necessary (using the same index
notation as Ref.~\cite{Ren:2011mh}) gives the operator structure
\begin{equation}
	\chi_{ijk} \chi_{k' mn} \chi^\dagger_{ijm} \Phi_{n'} \LC_{kk'} \LC_{nn'}
	+ \text{h.c.} 
	\label{}
\end{equation}
Using the Hilbert series has allowed the quick identification of an operator
that was missed in the original potential; a clear demonstration of the
practical efficacy of this technique.

The 31 independent dimension-6 operators for this example can also be easily
generated. The resulting operators are
\begin{equation}
\begin{gathered}
	3\; (\chi^\dagger \chi)^3, \quad
	2\; \Phi^\dagger (\chi^\dagger)^3 \chi^2, \quad
	2\; (\Phi^\dagger)^2 \chi (\chi^\dagger )^3, \quad
	(\Phi^\dagger)^3 (\chi^\dagger)^3, \quad
	2\; \Phi \chi^3 (\chi^\dagger)^2, \quad \\
	4\; \Phi \Phi^\dagger \chi^2 (\chi^\dagger)^2, \quad
	3\; \Phi (\Phi^\dagger)^2 \chi (\chi^\dagger)^2, \quad
	\Phi (\Phi^\dagger)^3 (\chi^\dagger)^2, \quad
	2\; \Phi^2 \chi^3 \chi^\dagger, \quad
	3\; \Phi^2 \Phi^\dagger \chi^2 \chi^\dagger, \quad \\
	3\; \Phi^2 (\Phi^\dagger)^2 \chi \chi^\dagger, \quad 
	\Phi^2 (\Phi^\dagger)^3 \chi^\dagger, \quad
	\Phi^3 \chi^3, \quad
	\Phi^3 \Phi^\dagger \chi^2, \quad
	\Phi^3 (\Phi^\dagger)^2 \chi, \quad
	(\Phi^\dagger \Phi )^3 .
\end{gathered}
\label{}
\end{equation}

\section{Thoughts on derivatives}
\label{deriv}

The examples we have worked through so far have been limited to collections of
spurions without derivatives. Derivatives are necessary if we wish to apply the
Hilbert series to a wider set of problems, such as the full set of SMEFT
operators of a given dimension. An immediate complication when including
derivatives are the equations of motion (EOM) -- relations among operators that
are not governed by symmetries or invariances.

At first glance, it seems like the Hilbert series is ill-equipped to handle EOM.
However, the Hilbert series is not a Lagrangian and is merely a tool to count
invariants formed from whatever spurions are put into the PE/PEF. For the
purposes of counting invariants, the role of the EOM is to remove spurions --
namely derivatives on fields -- by swapping them for different combinations of
spurions with no derivatives, which {\em is} something that can be handled by
the Hilbert series methodology.  Consider the EOM for the left-handed Standard
Model quark doublet $Q$:
\begin{equation}
	i \slashed{D}Q = y_u^\dagger \, u^{c \dagger}\epsilon H^* + y_d^\dagger
	\, d^{c \dagger} H,
\label{deriv_eom}
\end{equation}
which allows one to remove $\slashed D Q$ from the set of spurions. 
Derivatives of fermions can still appear in invariants, first showing up at
dimension 7~\cite{Lehman:2014jma}. For example, consider the operator
\begin{equation}
	\mathcal O_{LLud} = \LC_{ij}\, (d^c \sigma_\mu u^{c\dagger} ) (L^i D^\mu
	L^j ) .
\end{equation} 
The difference between the derivatives in the two equations above lies in the
Lorentz group representation.  Acting on a left-handed fermion with a derivative, we get
\begin{equation}
	\left(\frac{1}{2},\, \frac{1}{2}\right) \otimes \left(0,\,
	\frac{1}{2}\right) = \left(\frac{1}{2},\, 0\right)
	\oplus \left(\frac{1}{2},\, 1\right).
\end{equation}
The EOM in Eq.~(\ref{deriv_eom}) involves the $(\frac 1 2, 0)$ representation only.
Therefore we can incorporate
the EOM for a fermion field $\psi$ by including an additional spurion for the
$(\frac 1 2, 1)$ part of $D_{\mu}\psi$, but {\em omitting} the $(\frac 1 2, 0)$ spurion.

If we follow this logic, as we add further derivatives $D_{\mu}D_{\nu}\psi$,
etc.  we should add a new spurion to the PE/PEF for every new representation of
the Lorentz group that is formed.  For derivatives of scalars such as the Higgs
field $H$, $D_{\mu}H$ is new and should be added to the PE.  Further derivatives, such as
$D_{\mu}D_{\nu}H$, will contain a piece with $(0, 0)$ Lorentz
structure; this piece should not be included as a separate spurion as it is
eliminated by the Higgs EOM, $\Box H = m^2 H + \cdots$. Similarly, for
derivatives of the field strength tensors $D_{\lambda}X^{\pm}_{\mu\nu}$, the
$(\frac 1 2, \frac 1 2)$ representation should be omitted.  

Even after removing the derivative spurions, the Hilbert series still lacks
information about integration by parts, so invariants involving derivatives need
to be checked manually for redundancy. This check will get tedious if we Taylor
expand the Hilbert series to arbitrarily high mass dimension operators, but it
should be reasonably manageable for the dimension-8 operators.  Further
investigation along these lines will appear in Ref.~\cite{new}.

\section{Conclusion}
\label{conclusion}

The Hilbert series is a mathematical method providing a generating function for
group invariants. Since this method is not well known in the phenomenological
community, we have introduced the Hilbert series through simple examples and
provided several practical illustrations of its use. Through the presented
cases, we see that the Hilbert series technique proves very useful for common
applications. Any calculation where a BSM or effective field theory Lagrangian
contains fields transforming under uncommon group representations or where
operators of higher order in mass dimension are needed, the Hilbert series is an
easy way to get the right number of invariants. This is important in particular
when looking for small deviations from known physics by comparing to precision
measurements as redundant or missing terms in Lagrangians could invalidate
analyses.

With the inclusion of derivatives and the equations of motion, and applying
sufficient skill towards reducing the computational load of calculating the
necessary residues, it should be possible to compute a complete Hilbert series
for the Standard Model effective field theory. This generating function would
contain at each dimension a coefficient specifying the total number of
independent operators of that dimension. If the task was extended to calculate
the multi-graded Hilbert series, one would then be able to directly see the
different forms of operators at each order simply by expanding out the Hilbert
series. We intend to work towards this goal in an upcoming study \cite{new}.

\acknowledgments
Landon Lehman thanks the Galileo Galilei Institute for Theoretical Physics for
hospitality while some of this work was completed. The work of AM was partially
supported by the National Science Foundation under Grant No.~PHY-1417118. We
thank Aneesh Manohar for useful comments.

\appendix
\section{Mathematical background}
\label{basics}

\subsection{Characters}

The character of an irreducible group representation is the trace of the matrix
giving the representation. Any matrix representing a group element in a given
representation will have the same character.

In general, the characters of irreducible Lie group representations are obtained
from the Cartan matrices, using the Freudenthal recursion formula to find the
correct multiplicities of weights (see \cite{Hanany:2014dia} and references
therein). In practice, this calculation can be trivially done using the
Mathematica package LieART by utilizing the \texttt{WeightSystem} command
\cite{Feger:2012bs}. The results for some common representations are shown in
Table \ref{character_table}.  The characters can also be found using the
character generating functions outlined in \cite{Hanany:2014dia}.

\begin{table}[h]
\begin{center}
\begin{tabular}{cc}
\toprule
{\bf Representation} & {\bf Character function} \\
\midrule
$SU(2)$ fundamental & $ z + \frac{1}{z}$ \\
\midrule
$SU(2)$ quadruplet & $ z^3 + z + \frac{1}{z} + \frac{1}{z^3}$ \\
\midrule
$SU(3)$ fundamental & $ z_1 + \frac{z_2}{z_1} + \frac{1}{z_2}$ \\
\midrule
$SU(3)$ anti-fundamental & $\frac{1}{z_1} + \frac{z_1}{z_2} + z_2 $ \\
\midrule
$SU(3)$ adjoint & $ z_1 z_2 + \frac{z_2^2}{z_1}+\frac{z_1^2}{z_2}+ 2 +
\frac{z_2}{z_1^2} + \frac{z_1}{z_2^2} + \frac{1}{z_1 z_2}$ \\
\midrule
$SU(2)$ adjoint &  $ z^2 + 1 + \frac{1}{z^2}$ \\
\midrule
$U(1)$ with charge $Q$ & $ z^Q$ \\
\bottomrule
\end{tabular}
\caption{Character functions for several common group representations. Note that
setting all of the variables to 1 gives the dimension of the representation.}
\label{character_table}
\end{center}
\end{table}

Many common group calculations can be done using characters. For example,
consider taking the tensor product of two $SU(2)$ fundamentals. Using
the characters from Table \ref{character_table}, this looks like
\begin{equation}
	\left( z + \frac{1}{z} \right) \left( z + \frac{1}{z} \right) 
	= z^2 + 2 + \frac{1}{z^2} 
	= 1 + \left( z^2 + 1 + \frac{1}{z^2} \right), 
	\label{}
\end{equation}
which is just the familiar triplet-singlet spin decomposition $\frac{1}{2}
\otimes \frac{1}{2} = 0 \oplus 1$.

\subsection{The plethystic exponential}
\label{app:PE}

For a multivariable function $f(t_1, \dots ,t_n)$ satisfying the property of
going to zero at the origin, the plethystic exponential (PE)
\cite{Benvenuti:2006qr, Feng:2007ur, Hanany:2008sb, Gray:2008yu, Hanany:2014dia}
is
\begin{equation}
	\text{PE} [f(t_1,\dots,t_n)] \equiv \exp{ \left( \sum_{r=1}^{\infty}
	\frac{1}{r} f(t_1^r, \dots, t_n^r) \right)}.
	\label{}
\end{equation}
The plethystic exponential generates all symmetric combinations of the variables
of the function $f(t_1,\dots,t_n)$. For example, the plethystic exponential of
$f(A,B) = A + B$ is 
\begin{equation}
	\PE [ A + B ] =  \frac{1}{(1-A)(1-B)} = 1 + A + B + A^2 + A B + B^2 +
	\dots 
	\label{}
\end{equation}

For fermionic variables we are interested in antisymmetric combinations instead
of symmetric combinations, since fermions obey Fermi-Dirac statistics. The
fermionic plethystic exponential is defined as \cite{Hanany:2014dia}:
\begin{equation}
	\text{PEF} [ f(t_1,\dots, t_n) ] \equiv \exp \left( \sum_{r=1}^{\infty}
	\frac{ (-1)^{r+1}}{r} f(t_1^r,\dots , t_n^r ) \right) .
	\label{}
\end{equation}
For example, 
\begin{equation}
	\text{PEF} [ A + B ] = 1 + A + B + A B ,
	\label{}
\end{equation}
where the first three terms are trivial, and the last term is indeed
antisymmetric under $A \leftrightarrow B$, since fermions anticommute.  In the
body of this paper, the fermionic plethystic exponential is not always
explicitly differentiated from the ordinary plethystic exponential, but the
factor of $(-1)^{r+1}$ is always included with fermionic variables (spurions).

\subsection{Group integration with the Haar measure}
\label{haar_measure}

It is possible to integrate over the manifold of a Lie group by using an
invariant measure known as the Haar measure.  This group integration projects
out invariant quantities from the combinatorial expansion of characters provided
by the plethystic exponential.  The Haar measures that are used in this paper
are as follows (taken from \cite{Hanany:2008sb}):
\begin{equation}
	\int_{U(1)} \d \mu_{U(1)} = \frac{1}{2\pi \text{i}} \oint_{|z|=1|}
	\frac{\d z}{z} ,
	\label{}
\end{equation}

\begin{equation}
	\int_{SU(2)} \d \mu_{SU(2)} = \frac{1}{2\pi \text{i}} \oint_{|z|=1}
	\frac{\d z}{z} (1-z^2) ,
\label{haar_su2}
\end{equation}

\begin{equation}
	\int_{SU(3)} \d \mu_{SU(3)} = \frac{1}{(2\pi \text{i})^2} \oint_{|z_1|=1}
	\frac{\d z_1}{z_1} \oint_{|z_2|=1} \frac{\d z_2}{z_2} (1-z_1 z_2)
	\left(1-\frac{z_1^2}{z_2}\right)\left(1-\frac{z_2^2}{z_1}\right) .
\end{equation}
Further examples of Haar measures for various Lie groups can be found in
\cite{Hanany:2008sb}, where a general formula is also presented.

\bibliography{Notes}
\bibliographystyle{jhep}

\end{document}